\documentclass[12pt]{article}
\pdfoutput=1
\usepackage{putex}
\usepackage{amsmath}
\usepackage{amssymb}
\usepackage{simpler-wick}
\usepackage{graphicx}
\usepackage{epstopdf}
\usepackage{enumerate}
\usepackage{cite}
\usepackage{tensor}
\usepackage{slashed}
\usepackage{feynmf}
\usepackage{hyperref}
\usepackage{bbold}
\usepackage{mathrsfs}
\usepackage{caption}
\usepackage{subcaption}
\usepackage{axodraw2}
\usepackage{pstricks}
\usepackage{color}
\usepackage{youngtab}
\usepackage{amsfonts}
\usepackage{braket}

\newcommand\numberthis{\addtocounter{equation}{1}\tag{\theequation}}
\def\O{{\mathcal O}}

\usepackage[utf8x]{inputenc}
\usepackage{braket}
\usepackage{titlesec}
\usepackage{fixltx2e}
\usepackage{cleveref}
\hypersetup{breaklinks}
\usepackage{comment}

\newcommand {\be} {\begin {equation}}
\newcommand {\ee} {\end {equation}}
\newcommand{\bea}{\begin{eqnarray}}
\newcommand{\eea}{\end{eqnarray}}

\makeatletter
\renewcommand{\@maketitle}{
	\newpage
	\begin{center}%
		{\large\bfseries \@title \par}%
	\end{center}%
	\par} \makeatother

\numberwithin{equation}{section}

\titleformat*{\section}{\large\bfseries}

\institution{KY}{Department of Physics and Astronomy, University of Kentucky, Lexington, KY, 40506}

\begin{document}

	\authors{Hardik Bohra\worksat{\KY}
, Ashish Kakkar\worksat{\KY}
, Allic Sivaramakrishnan\worksat{\KY} 
\let\thefootnote\relax\footnote{\texttt{${}$bohra.hardik@uky.edu }}
\let\thefootnote\relax\footnote{\texttt{${}$aka283@g.uky.edu}}
\let\thefootnote\relax\footnote{\texttt{${}$allicsiva@uky.edu}}
 }
 
	\title{Information Geometry and Holographic Correlators}
	
	\abstract{
		We explore perturbative corrections to quantum information geometry. In particular, we study a Bures information metric naturally associated with the correlation functions of a conformal field theory. We compute the metric of holographic four-point functions and include corrections generated by tree Witten diagrams in the bulk. In this setting, we translate properties of correlators into the language of information geometry. Cross terms in the information metric encode non-identity operators in the OPE. We find that the information metric is asymptotically AdS. Finally, we discuss an information metric for transition amplitudes.
			}
	
	\date{}
	
	\maketitle
	\setcounter{tocdepth}{2}
	\tableofcontents
	
	\section{Introduction}
	
How does quantum information encode effective field theory? This question is relevant in holography, where the quantum extremal surface proposal for quantum corrections implies novel features of black hole evaporation \cite{Engelhardt:2014gca, Penington:2019npb,Almheiri:2019psf}. Despite recent progress, effective field theory remains far less understood in terms of quantum information than in the language of Lagrangians, correlation functions, and the S-matrix. Developing this subject may prove useful. We may learn more about effective field theory via constraints coming from quantum information. We may also identify new perturbative structures in a quantum information description of gravity. Even a better technical understanding of quantum corrections at first order may have far-reaching implications for our understanding of black holes.

While computations of entanglement entropy in AdS/CFT have been illuminating, the mechanics of effective field theory can be studied in another setting as well. In certain cases, quantum information quantities can be related directly to correlation functions or the S-matrix, or to their ingredients. For recent results in this direction, see for example \cite{Balakrishnan:2020lbp,Balakrishnan:2019gxl,Faulkner:2016mzt,May:2019odp,Belin:2021htw,Bose:2020shm,Jordan:2011ci}). This approach exposes the role of effective field theory, allowing direct study of its interplay with quantum information.
	
	In this work we study the Bures information metric, which is a measure of the distinguishability of nearby states. We explore perturbative corrections and focus in particular on the information metric associated with correlators in holographic conformal field theories (CFTs). To summarize the setup, we consider the Bures distance between pure states $D_B(\psi(x_1,x_2),\psi(x_3,x_4))^2$ near $x_1 = x_3, x_2 = x_4$ using states,
	\begin{equation}
	\ket{\psi(x_1,x_2)} = \frac{\O_2(x_2)\O_1(x_1)\ket{0} }{\sqrt{\braket{\O_1(x_1^*) \O_2(x_2^*)  \O_2(x_2)\O_1(x_1) }}},
	\end{equation}
	where we use the notation $(\O(x))^\dagger = \O(x^*)$. Up to a normalization factor, the Bures distance is a four-point function,
	\begin{equation}
	D_B(\psi(x_1,x_2),\psi(x_3,x_4))^2 \sim \braket{\O_1 (x_1^*)\O_2 (x_2^*)\O_1(x_3)\O_2(x_4)}.
	\end{equation}
	In a holographic CFTs, the Bures information metric of this two-operator state is
	\begin{equation}
	g_{x_1^\mu x_2^\nu} \equiv \frac{d^2}{dx_1^\mu dx_2^\nu} D_B(\psi(x_1,x_2),\psi(x_3,x_4))^2 \approx g_{x_1^\mu x_2^\nu}^{(0)} + \frac{1}{N^2} g^{(2)}_{x_1^\mu x_2^\nu} + \ldots
	\label{InfoMetricIntro}
	\end{equation}
	and encodes features of four-point correlators in a simple way.\footnote{To highlight the physics involved, we will refer to this metric as the metric of the correlator. Note that there is a one-to-one mapping between the information distance of two $n$-operator states and a certain set of $2n$-point correlators in the appropriate kinematic configuration. The normalization factor is understood, though note that it is a correlator as well. A ratio of correlators may seem strange, but universal properties will appear nevertheless.}  Taking a similar approach, we also discuss transition amplitudes induced by a unitary $U = e^{-i \lambda H}$ and work perturbatively in $\lambda$. Our aim here is to take initial steps in describing the information geometry of $2n$-point processes in quantum field theory, but it is straightforward to explore this story more fully using standard methods.
	
	As four-point functions appear explicitly, the connection to $1/N$ perturbation theory is direct. The $1/N$ corrections are computed by four-point Witten diagrams in the bulk, which have been studied extensively at tree level and more recently at one loop \cite{MeltzerPS19,MeltzerS20,AldayC17,AharonyABP16,Bissi:2020woe,Alday:2021ajh,Carmi:2021dsn,Carmi:2019ocp}. By using these known results as input, computing the information metric itself is relatively simple \eqref{InfoMetricIntro}. This approach applies equally well in all dimensions. By comparison, it is more challenging to probe $1/N$ corrections by taking the partial trace of density matrix \cite{Kusuki:2019hcg,Suzuki:2019xdq}. The reduced density matrix approach probes entanglement wedge structure and has been explored to order $\mathcal{O}(N^0)$ in CFT$_2$. However, we expect that computing $1/N$ corrections on a replica manifold will be more challenging than in the original theory, and far less tractable in general dimensions. In short, the pure state and reduced density matrix approaches probe different and complementary features of the information metric, and may be useful for different purposes.
	
	Here, we give an outline of this paper. In Section 2, we review basics of information geometry and then discuss perturbative corrections to the information metric. We show that when the fidelity factorizes, the information metric also factorizes. In Section 3, we review the two-point function metric derived in \cite{Kusuki:2019hcg,Suzuki:2019xdq}, and then study the metric of four-point functions in CFT$_d$. We show that correlators with a weak-coupling expansion have an information metric that also has a weak-coupling expansion. In Section 4, we compute the information metric in explicit four-point examples. Correlators in mean field theory (MFT) with pairwise-identical operators factorize and have a factorized information metric. MFT correlators with identical scalars do not factorize, or equivalently have operators besides the identity exchanged in every channel. The resulting information metric does not factorize. We then compute an $\mathcal{O}(1/N^2)$ correction dual to a tree Witten diagram in bulk $\phi_1^2 \phi_2^2$  theory. While the MFT contribution factorizes, the tree-level information metric does not, and the tree diagram has operator exchanges besides the identity in all channels. In all four-point examples, we find that the information metric is asymptotically AdS. In Section 5, we address similar questions for transition amplitudes of qubits, a simple model for the S-matrix. In this context, we find an information metric for transition amplitudes with identical in and out states. The metric takes the form $\braket{H^2}-\braket{H}^2$. In Section 6, we discuss future directions.

	\section{Information metric basics}
	
	\subsection{Review}
	We review the Bures distance $D_B^2$ and the associated metric, which we will refer to as the information metric. We follow the approach in \cite{Suzuki:2019xdq,Kusuki:2019hcg}, to which we refer the reader for further details and discussion of other distance measures. The Bures distance between density matrices $\rho_1,\rho_2$ is
	\begin{equation}
	D_B(\rho_1,\rho_2)^2 = 2\left(1-\sqrt{F(\rho_1,\rho_2)}\right),
	\end{equation}
	where $F$ is the fidelity,
	\begin{equation}
	F(\rho_1,\rho_2) = \left( \text{tr} \left( \sqrt{ \sqrt{\rho_1} \rho_2 \sqrt{\rho_1}    }\right) \right)^2.
	\end{equation}
	Though not manifest above, fidelity is symmetric. We will study pure states, for which $D_B^2$ takes a simple form. When $\rho_i = \ket{\psi_i}\bra{\psi_i}$ with $\ket{\psi_i}$ normalized, $F(\rho_1,\rho_2) = |\braket{\psi_1| \psi_2}|^2$ and
	\begin{equation}
	D_B(\rho_1,\rho_2)^2 = 2(1-|\braket{\psi_1| \psi_2}| ).
	\end{equation}
	In other words, the Bures distance between pure states is simply the magnitude of the inner product. 
	
	Consider a family of density matrices $\rho(\lambda_i)$ that depend smoothly on parameters $\lambda_i$. The Bures distance of nearby $\rho$'s can be described by a metric as
\begin{equation}
	D_B(\rho (\lambda_i ), \rho(\lambda_i+d\lambda_i ))^2 \approx \sum_i d\lambda_i \frac{d}{d\lambda_i' }\bigg|_{\lambda_i' = \lambda_i} D_B(\rho (\lambda_i ), \rho(\lambda_i'))^2   +\sum_{i,j} g_{i j}(\lambda_i) d\lambda_i d\lambda_j,
	\end{equation}
	where
	\begin{equation}
	g_{i j}  \equiv \frac{d^2}{d\lambda_i' d\lambda_j'} \bigg|_{\substack{\lambda_i' = \lambda_i\\ \lambda_j' = \lambda_j}} \sqrt{F(\rho(\lambda_i), \rho(\lambda_i'))}
	\end{equation}
	is the Bures metric. Assuming $D_B(\rho(\lambda_i),\rho(\lambda_i'))^2$ is analytic in a neighborhood of $\lambda_i' = \lambda_i$, then it has a minimum at $\lambda_i' = \lambda_i$ and so
	\begin{equation}
		D_B(\rho (\lambda_i ), \rho(\lambda_i+d\lambda_i ))^2 \approx \sum_{i,j} g_{i j}(\lambda_i) d\lambda_i d\lambda_j,
	\end{equation}
	The information metric therefore captures the distinguishability of nearby density matrices. Following the quantum Cramer-Rao bound, the inverse metric $g_{ij}^{-1}$ bounds the error in estimating values of $\lambda_i$ through measurement.
	
	\subsection{Perturbative corrections}
	
	We now study the information metric in the context of perturbation theory. For a family of density matrices $\rho$ parametrized by $\lambda_1, \lambda_2$, 
	\begin{equation}
	D_B(\rho(\lambda_1,\lambda_2), \rho(\lambda_1,\lambda_2+d\lambda_2) )^2 \approx g_{2 2}(\lambda_1,\lambda_2) (d\lambda_2)^2.
	\end{equation}
	Suppose $\rho(\lambda_1,\lambda_2)$ has an expansion in $\lambda_1$ about for example $\lambda_1 = \lambda$. It follows that $g_{22}(\lambda_1,\lambda_2)$ can also be expanded in $\lambda_1$,
	\begin{equation}
	g_{2 2}(\lambda_1,\lambda_2)  = \sum_{n=0} g_{2 2}^{(n)}(\lambda,\lambda_2) (\lambda_1-\lambda)^n.
	\label{generalmetricexpansion}
	\end{equation}
	This statement is intuitive when the $g_{2 2}^{(n)}(\lambda_1,\lambda)$ are computed from objects within the same Hilbert space, which is natural in quantum mechanics. In weakly coupled quantum field theory, expanding an interacting quantity in a coupling $\lambda_1$ gives $g^{(n)}_{22}(\lambda,\lambda_2)$ computed from elements of the Hilbert space of the free theory. $\lambda_2$ parametrizes the state in the exact theory. Concretely, when $\lambda_1$ is a coupling constant, $\lambda_2$ can be the position or momentum that specifies the state.
		
	Finally, we show that factorization of the fidelity into the fidelities of subsystems implies factorization of the information metric. Suppose that
	\begin{equation}
	D_B(\rho(\lambda_1,\lambda_2) ,\rho(\lambda_3 ,\lambda_4))^2 = 2\left(1-\sqrt{F_1(\lambda_1,\lambda_3) F_2(\lambda_2, \lambda_4)} \right),
	\end{equation}
	where $F_1, F_2$ are themselves fidelities, 	
	\begin{align*}
	D_B(\rho_1 (\lambda_1),  \rho_1 (\lambda_3)) ^2&= 2\left(1-\sqrt{F_1(\lambda_1,\lambda_3)}\right), 
	\\
	D_B(\rho_2 (\lambda_2),  \rho_2 (\lambda_4)) ^2&= 2\left(1-\sqrt{F_2(\lambda_2,\lambda_4)}\right),
	\numberthis
	\end{align*}
	for families of density matrices $\rho_1(\lambda_1), \rho_2(\lambda_2)$ that admit information metrics $g_{1 1} d\lambda_1^2 $ and $ g_{2 2}  d \lambda_2^2$ respectively. Expanding $D_B(\rho(\lambda_1,\lambda_2) ,\rho(\lambda_3 ,\lambda_4))^2$ using $\lambda_3 = \lambda_1 + d\lambda_1$ and $\lambda_4 = \lambda_2 + d\lambda_2$ therefore gives what we refer to as a factorized metric,
	\begin{equation}
	D_B(\rho(\lambda_1,\lambda_2), \rho(\lambda_1 + d\lambda_1,\lambda_2 + d\lambda_2)) ^2\approx g_{\lambda_1 \lambda_1} d\lambda_1^2 + g_{\lambda_2 \lambda_2} d\lambda_2^2.
	\label{MetricFactorization}
	\end{equation}
	The cross term
	\begin{align*}
	\frac{d^2}{d\lambda_3 d\lambda_4 }\bigg |_{\substack{\lambda_3 = \lambda_1 \\ \lambda_4 = \lambda_2}} &D_B(\rho(\lambda_1,\lambda_2) ,\rho(\lambda_3 ,\lambda_4))^2 
	\\
	&=
	-2\left(
	\frac{d}{d\lambda_3} \bigg |_{\lambda_3 = \lambda_1} \sqrt{F(\rho_1(\lambda_1),\rho_1(\lambda_3))} \right)
	\left(
	\frac{d}{d\lambda_4} \bigg |_{\lambda_4 = \lambda_2} \sqrt{F(\rho_2(\lambda_2),\rho_2 (\lambda_4))} 
	\right)
	=0,
	\numberthis
	\end{align*}
	because each factor is the first order term in $D_B(\rho_1 (\lambda_1),  \rho_1 (\lambda_1 + d\lambda_1))^2$ and $ D_B(\rho_2 (\lambda_2) \rho_2 (\lambda_2 + d\lambda_2))^2$ respectively. As these Bures distances admit information metrics by assumption, the first order terms are zero. An immediate corollary is that the presence of cross terms in the metric implies the failure of factorization of fidelity into sub-fidelities.\footnote{While it may be true that the absence of cross terms implies factorization into sub-fidelities, we do not claim this. In principle, the fidelity could factorize into two functions that are not themselves fidelities.} As we will see shortly, this notion of factorization will be related to factorization in correlators.
		
	\section{CFT correlators}
	
	Our main focus will be four-point correlators in holographic CFT$_d$. Some of the explicit expressions we give will be for CFT$_2$ for simplicity. Nevertheless, we expect many of our conclusions apply more generally.
	
	\subsection{Review: two-point function}
	
	Following \cite{Suzuki:2019xdq,Kusuki:2019hcg}, we review the information metric for the Euclidean two-point function of scalar primaries. We work with real operators, which obey $(\mathcal{O}(x,\tau))^\dagger =  \mathcal{O}(x,-\tau)$ \cite{Simmons-Duffin16TASI}. We begin with density matrix
	\begin{equation}
	\rho(x) = \frac{\O(x)\ket{0} \bra{0} \O (x^*)}{\braket{\O(x) \O(x^*)}},
	\end{equation}
	where $\frac{\mathcal{O}(x)\ket{0}}{\sqrt{\braket{\O(x^*)\O(x) }}}$ has unit norm. We use notation $x^\mu = (x^i,\tau)$ with $(x^\mu)^* \equiv (x^i,-\tau)$ and suppress the indices in the arguments of $\O$ for compactness. Raised indices run over spatial coordinates while subscripts label external points. Expectation values are taken in the vacuum. The information distance is
	\begin{equation}
	D_B(\rho(x_1),\rho(x_2))^2 = 2\left( 1-\frac{|\braket{\O (x_1^*)\O(x_2)}|}{\sqrt{\braket{\O (x_1^*) \O(x_1)}\braket{\O  (x_2^*) \O(x_2)}}} \right).
	\end{equation}
	The CFT two-point function is fixed by conformal symmetry to be $\braket{\mathcal{O}(x) \mathcal{O}(y)} = (x-y)^{-2\Delta}$, where $\Delta$ is the scaling dimension of $\O$.
	\begin{equation}
	D_B(\rho(x_1),\rho(x_2))^2 = 2\left( 1-\frac{(4\tau_1 \tau_2)^{\Delta}}{((x_1^i-x_2^i)^2+(\tau_1+\tau_2)^2)^\Delta} \right).
	\end{equation}
	An information metric is obtained from the expansion
	\begin{equation}
	x_2^\mu = x_1^\mu + dx^\mu_1.
	\end{equation}
	The resulting metric describes the distinguishability of states created by inserting operators at nearby locations.	The information metric in CFT$_2$ is \cite{Suzuki:2019xdq, Kusuki:2019hcg}
	\begin{equation}
	ds^2 = \frac{\Delta}{2\tau_1^2}\left(dx_1^2+d\tau_1^2 \right),
	\end{equation}
	which is proportional to the metric of Poincare AdS$_3$. See \cite{Kusuki:2019hcg} for additional examples of this equivalence. The general dimension case is similar to the two-dimensional case. For relating CFT$_2$ expressions to those in CFT$_d$, it is useful to note that $\frac{d}{dx_1^i} \frac{d}{dx_1^j}((x_1^i-x_2^i)^2+(\tau_1+\tau_2)^2) = 2\frac{d}{dx_1^j} (x_1-x_2)^i = 2\delta^{ij}$. This implies $g_{x^i x^j} \sim \delta^{ij}$ for the two-point function metric. As we expand about $x_2^i = x_1^i$, we also have $g_{x^i \tau} = g_{\tau x^i } = 0$. The information metric in $d$-dimensions is therefore
	\begin{equation}
	ds^2 = \frac{\Delta}{2\tau^2_1}\left(\sum_i(dx^i_1) ^2+d\tau^2_1 \right),
	\end{equation}
	which is proportional to the Euclidean Poincare AdS$_d$ metric. As the two-point function is determined by conformal symmetry, this metric is the same for all CFT$_d$. The reduced density matrix obtained by tracing out a spatial subregion does probe theory-dependent information \cite{Suzuki:2019xdq, Kusuki:2019hcg}, but we take a different approach here.
	
	\subsection{The four-point function}
	
	In order to obtain a theory-specific information metric, we now turn to two-operator states,
	\begin{equation}
	\rho(x_1,x_2) = \frac{\O_2(x_2)\O_1(x_1)\ket{0} \bra{0}\O_1(x_1^*)\O_2(x_2^*)}{\braket{\O_1(x_1^*) \O_2(x_2^*)  \O_2(x_2)\O_1(x_1) }}.
	\end{equation}
	The Bures distance is
	\begin{equation}
	D_B(\rho(x_1,x_2),\rho(x_3,x_4))^2 = 2\left( 1-\frac{|\braket{\O_1(x_3^*)\O_2(x_4^*)\O_2 (x_2 )\O_1 (x_1)}|}{\sqrt{
			\braket{\O_1 (x_1^*)\O_2 (x_2^*)\O_2(x_2)\O_1(x_1)}
			\braket{\O_1 (x_3^*) \O_2 (x_4^*) \O_2(x_4)\O_1(x_3)}
		} 
	}
	\right).
	\end{equation}
	With $\tau_1 < \tau_2 < 0 < -\tau_4 < - \tau_3$, the correlators above are time-ordered in Euclidean.\footnote{See \cite{Simmons-Duffin16TASI} for discussion.} This expression is a valid Bures distance for all $\tau_i < 0$.
	
	Various limits of $D_B^2$ are determined by familiar properties of the four-point function. $D_B^2$ is finite in the OPE limits $x_{12}^2 \rightarrow 0, x_{34}^2 \rightarrow 0$ and determined by the $\mathcal{O}_1 \mathcal{O}_2$ OPE. In the limit $\tau_i \rightarrow 0$, the normalization factor diverges and gives $D_B^2 \rightarrow 0$. As is standard, $\tau$ acts as a UV regulator for a state formed by inserting local operators, which would otherwise contain arbitrarily high energy excitations. Cluster decomposition implies that when we translate $x_3, x_4$ by a large distance, 
	\begin{equation}
	D_B(\rho(x_1,x_2),\rho(x_3,x_4))^2 \approx 2\left( 1-\frac{|\braket{\O_1 (x_1^*)\O_2 (x_2^*)}\braket{\O_1(x_3)\O_2(x_4)}|}{\sqrt{
			\braket{\O_1 (x_1^*)\O_2 (x_2^*)\O_2(x_2)\O_1(x_1)}
			\braket{\O_2 (x_4^*)\O_1 (x_3^*) \O_1(x_3)\O_2(x_4)}
		} 
	}
	\right).
	\end{equation}
	We consider the information metric obtained from the expansion
	\begin{equation}
	x_3^\mu = x_1^\mu + dx_1^\mu,
	\quad \quad \quad
	x_4^\mu = x_2^\mu + dx_2^\mu.
	\label{fourpointexpansion}
	\end{equation}
	One can check that the first-order terms are automatically zero,
		\begin{equation}
		\frac{d}{d x_3^\mu}\bigg |_{\substack{x_3 = x_1\\ x_4 = x_2}}D_B(\rho(x_1,x_2),\rho(x_3,x_4))^2 = 0 , ~~~~~~~\frac{d}{d x_4^\mu}\bigg |_{\substack{x_3 = x_1\\ x_4 = x_2}} D_B(\rho(x_1,x_2),\rho(x_3,x_4))^2 = 0.
		\end{equation}
		In the next section, we find that the small-$\tau$ limit gives Euclidean Poincare AdS, 
		\begin{equation}
		\tau_k \rightarrow 0: ~~~~~~~~~~~~~~~~
		 g_{\mu \nu}dx^\mu  dx^\nu \approx  \frac{\Delta_k}{2}  \frac{\sum_i (dx_k^i)^2+d\tau_k^2}{\tau_k^2}
		.
		\end{equation}
		Specifically, we will show that the identity contribution to the OPE in the $13 \rightarrow 24$ channel gives an asymptotically-AdS information metric. We expect this is the leading contribution to the information metric for general correlators, including at higher points.
				
	As the four-point function is theory-dependent, we can study perturbative corrections. Suppose the states have a perturbative expansion in $\lambda_1$ about $\lambda$. The correlators and $D_B^2$ can also be expanded in $\lambda_1$. To every order in $\lambda_1-\lambda$, the Bures distance is 0 for $x_3 = x_1, x_2 = x_4$ because the states are identical at these locations by construction for all $\lambda_1$. $x_3 = x_1, x_2 = x_4$ is therefore a minimum of $D_B^2$ for any $\lambda_1$, and so the information metric that arises from expanding about this point is still the leading correction to $D_B^2$ at all orders in $\lambda_1-\lambda$. \eqref{generalmetricexpansion} now follows, which is that the metric has an expansion to all orders in $\lambda_1-\lambda$:
	\begin{equation}
	g_{\mu \nu} = \frac{d^2}{d x_3^\mu dx_4^\nu}\bigg |_{\substack{x_3 = x_1\\ x_4 = x_2}}  D_B(\rho(x_1,x_2,\lambda_1),\rho(x_3,x_4,\lambda_1))^2 
	=
	\sum_{n = 0}  g^{(n)}_{\mu \nu}(\lambda) (\lambda_1-\lambda)^n.
	\label{FourPtMetric}
	\end{equation}
	The same argument applies to states created by $n$ operator insertions.
	
	\section{Four-point examples}
	
	We now demonstrate the statements in the previous section. We first study the MFT correlator, which captures the contribution of the identity operator to any CFT correlator. We find the MFT information metric is asymptotically AdS. We then consider a $1/N^2$ correction in holographic CFTs computed by a tree Witten diagram in the bulk. We find that the tree-level contribution preserves the asymptotically AdS behavior of the information metric, consistent with the fact that the block decomposition of tree diagrams does not contain the identity exchange. We will often present CFT$_2$ expressions for notational simplicity.
	
	\subsection{Mean Field Theory}
	The MFT correlator is computed by taking Wick contractions as in free field theory. The MFT four-point function is
	\begin{equation}
	\braket{\O_1(x_1)\O_2(x_2)\O_3(x_3)\O_4(x_4)}_{MFT}
	=
	\frac{\delta_{\O_1,\O_2}\delta_{\O_3,\O_4}}{x_{12}^{2\Delta_1}x_{34}^{2\Delta_3}}
	+\frac{\delta_{\O_1,\O_3}\delta_{\O_2,\O_4}}{x_{13}^{2\Delta_1}x_{24}^{2\Delta_4}}
	+\frac{\delta_{\O_1,\O_4}\delta_{\O_3,\O_2}}{x_{14}^{2\Delta_1}x_{32}^{2\Delta_2}}
	.
	\end{equation}
	Choosing $\O_3=\O_1, \O_4=\O_2$  with $\O_1 \neq \O_2$, 
	\begin{equation}
	\braket{\O_1(x_1)\O_2(x_2)\O_1(x_3)\O_2(x_4)}_{MFT}
	=	x_{13}^{-2\Delta_1}x_{24}^{-2\Delta_2}.
	\end{equation}
	The Bures distance is
	\begin{equation}
	D_B(\rho(x_1,x_2),\rho(x_3,x_4))^2 
	=2\left(1-
	\frac{(2\tau_1)^{\Delta_1} (2\tau_2)^{\Delta_2}(2\tau_3)^{\Delta_1}(2\tau_4)^{\Delta_2}}
	{((x_1-x_3)^2+(\tau_1+\tau_3)^2 )^{\Delta_1}((x_2-x_4)^2+(\tau_2+\tau_4)^2 )^{\Delta_2}}
	\right).
	\end{equation}
	Using the expansion
	\begin{equation}
	x^\mu_3 = x^\mu_1+dx^\mu_1,\quad \quad \quad\quad\quad 
	x^\mu_4 = x^\mu_2+dx^\mu_2,
	\label{abc}
	\end{equation}
	the information metric for CFT$_d$ is
	\begin{equation}
	ds^2 = \frac{\Delta_1}{2\tau_1^2}\left( \sum_i (dx_1^i)^2 + d\tau_1^2\right) + \frac{\Delta_2}{2
	\tau_2^2}\left( \sum_i (dx_2^i)^2 + d\tau_2^2\right).
	\end{equation}
	The lack of cross terms in the metric above is consistent with \eqref{MetricFactorization}, as this MFT correlator factorized into products of lower-point correlators. In CFT language, this factorization is the statement that only the identity operator is exchanged in the $13 \rightarrow 24$ channel. Operators with the dimensions of double trace operators $[\O_1 \O_2]$ are exchanged in other channels.
	
	Next, we consider a MFT correlator with four identical operators, 
	\begin{equation}
	\braket{\O(x_1)\O(x_2)\O(x_3)\O(x_4)}_{MFT}
	=
	\frac{1}{x_{12}^{2\Delta}x_{34}^{2\Delta}}
	+\frac{1}{x_{13}^{2\Delta}x_{24}^{2\Delta}}
	+\frac{1}{x_{14}^{2\Delta}x_{32}^{2\Delta}}.
	\end{equation}
	Unlike the pairwise-identical case, this correlator does not factorize. The expression for the full metric is large but straightforward to obtain, so we only give some explicit expressions at specific values of $\Delta$. We have also checked that the first order terms vanish, confirming the information metric is the leading contribution to $D_B^2$. Even with $\Delta=1$ the full metric is a large expression, but it simplifies for $x_1 = x_2$, 
	\begin{align*}
	ds^2  = &\frac{\left(\tau _1-\tau _2\right){}^2}{2 \tau _1^2 \tau _2^2 \left(\tau _1+\tau _2\right){}^2 \left(\tau _1^4+14 \tau _2^2 \tau _1^2+\tau _2^4\right){}^2}
	\\
	&
	\bigg(
	( d\tau_1^2 \tau _2^2 +d\tau_2^2 \tau _1^2) \left(\tau _1^4+62 \tau _2^2 \tau _1^2+\tau _2^4\right) \left(\tau _1+\tau _2\right){}^4
	+(\tau _2^2dx_1^2 + \tau _1^2dx_2^2 ) \left(\tau _1+\tau _2\right){}^4  \left(\tau _1^2-\tau _2^2\right){}^2
	\\
	&
	+64 \left(dx_1 dx_2 \tau _1^4 \tau _2^4\left(\tau _1-\tau _2\right){}^2  
	- d\tau_1 d\tau_2 \tau _1^3 \tau _2^3 \left(\tau _1^4+3 \tau _2 \tau _1^3+8 \tau _2^2 \tau _1^2+3 \tau _2^3 \tau _1+\tau _2^4\right)\right)
\bigg)
	\numberthis
	\end{align*}
	The full metric (with $x_1 \neq x_2$) has the limiting behavior
	\begin{equation}
	\tau_1 \rightarrow 0: ~~~~~ ds^2 \approx \frac{1}{2\tau_1^2}(dx_1^2+d\tau_1^2) ,
	\end{equation}
	and similarly for $\tau_2 \rightarrow 0$ due to symmetry in $\tau_1, \tau_2$. For general $\Delta$, we find 
	\begin{equation}
	\tau_i \rightarrow 0: ~~~~~ds^2 \approx \frac{\Delta}{2\tau_i^2}(dx_i^2+d\tau_i^2),
	\end{equation}
	which we verified explicitly in $d=2$ up to $\Delta = 7$ for integer values of $\Delta$. As $x_i$ does not appear in the CFT$_2$ expression above, we expect the same asymptotic behavior in CFT$_d$.
	
	The information metric in the identical operator case did not factorize. In the correlator, operators above the identity are exchanged in the $13 \rightarrow 24$ channel. All cross terms in the metric associated with an OPE channel are therefore proportional to the OPE coefficients of some operator exchanged in that channel.
	
	\subsection{Holographic correction: tree level}
	Now we specialize to a holographic CFT. At each order in the $1/N$ expansion, correlators of light single trace operators are computed by Witten diagrams in AdS.\footnote{This is true only under certain assumptions and has only been studied in generality up to one loop, but these details will not be relevant in this work.} The $\mathcal{O}(N^0)$ contribution is dual to free propagation in the bulk and is computed by MFT correlators. The next correction to four-point functions occurs at $\mathcal{O}(1/N^2)$, and is computed by tree Witten diagrams. We consider pairwise identical operators $\O_1 = \O_3, \O_2 = \O_4$. We assume the bulk theory has a $(\phi_1\phi_2)^2$ vertex, where $\phi_i$ are dual to $\O_i$. The tree-level contribution is therefore the contact diagram
	\begin{equation}
	\mathcal{A}^{\phi_1^2 \phi_2^2}(x_i) 
	=
	\int_{AdS} d^{d+1}y \sqrt{-g} \prod_{i}^4 K_{\Delta_i}(x_i,y)
	\equiv 
	D_{\Delta_1 \Delta_2 \Delta_1 \Delta_2},
	\end{equation}
	where $K_{\Delta_i}(x,y)$ is the bulk to boundary propagator for the bulk field with boundary dual $\mathcal{O}_i$. For particular scaling dimensions, $D$-functions are known in closed form. For instance,
	\begin{equation}
	\frac{2x_{13}^2 x_{24}^2}{\Gamma\left( 2-\frac{d}{2}\right)} D_{1111}(x_i)
	=
	\frac{1}{z-\bar{z}} \left( 2 \text{Li}_2(z) - 2\text{Li}_2(\bar{z}) + \log(z \bar{z}) \log \frac{1-z}{1-\bar{z}}\right).
	\end{equation}
	The contact diagram for other integer scaling dimensions can be found using $D$-function identities. We choose $\O_1, \O_2$ to be distinct scalars with equal dimension, $\Delta_1 = \Delta_2 = 1$. We will once again work in $d=2$, in which $\Delta = 1$ is above the unitarity bound.
	The information metric is found by expanding $D_B^2$ in the small parameter $1/N^2$. We have checked explicitly that the first order terms, $\frac{1}{N^2} dx_i^\mu$, are zero. The leading contribution to $D_B^2$ therefore comes from the information metric. The metric is
	\begin{equation}
	ds^2 = \frac{dx_1^2 + d\tau_1^2}{2\tau_1^2}+\frac{dx_2^2 + d\tau_2^2}{2\tau_2^2} + \sum_{i,j}\frac{1}{N^2} g^{(2)}_{i j }dx^i dx^j,
	\end{equation}
	where the leading term is the metric of pairwise identical MFT correlator studied earlier. The leading term factorizes but the $1/N^2$ correction does not. As in the MFT case, the explicit form of the $1/N^2$ contribution to the metric is lengthy. With $x_1=x_2$, the metric takes a simpler form,
	\begin{align*}
	&ds^2=
	\frac{\pi ^{d/2} \Gamma \left(2-\frac{d}{2}\right) }{\left(\tau _1-\tau _2\right){}^4 \left(\tau _1+\tau _2\right){}^4}
	\\
	&
	\bigg(-8 \left(\tau _1+\tau _2\right){}^2 \left(\tau _1-\tau _2\right){}^2 
	\left(\tau _1^2 ( dx_2^2+d\tau _2^2) -\tau _2 \tau _1(d\tau_1 d\tau_2+dx_1 dx_2) +\tau _2^2 \left(d\tau _1^2+dx_1^2\right)\right)
	\\
	&
	+\left(\tau _1-\tau _2\right){}^4 \log \left(\frac{\left(\tau _1-\tau _2\right){}^4}{16 \tau _1^2 \tau _2^2}\right) X_-(\tau_1,\tau_2)
	-2 \left(\tau _1+\tau _2\right){}^4 \log \left(\frac{\left(\tau _1+\tau _2\right){}^2}{4 \tau _1 \tau _2}\right) X_+(\tau_1,\tau_2)
	\bigg)
	\end{align*}
	where
	\begin{align*}
	X_-(\tau_1,\tau_2)&=
4 (\tau _1^2 (dx_2^2 +d\tau_2^2) + \tau _2^2(dx_1^2+d\tau _1^2 ))+(dx_1 dx_2 + d\tau_1 d\tau_2)\left(\tau _1^2-6 \tau _2 \tau _1+\tau _2^2\right),
	\\
	X_+(\tau_1,\tau_2)&=
-	4(\tau _1^2 (dx_2^2 +d\tau _2^2) +\tau _2^2( dx_1^2+d\tau _1^2  ))+(dx_1 dx_2 +d\tau _1 d\tau_2 )\left(\tau _1+\tau _2\right){}^2.
	\numberthis
	\end{align*}
	We have checked numerically that the full metric ($x_1 \neq x_2$) obeys 
	\begin{equation}
	\lim_{\tau_i \rightarrow 0}\tau_i ~ g_{\mu \nu}^{(2)}(\tau_1,\tau_2,x_1,x_2) = 0.
	\end{equation}
	In other words, $g_{\mu \nu}^{(2)}(\tau_1,\tau_2,x_1,x_2)$ does not change the $1/\tau_i^2$ divergence we found coming from the MFT contribution. The metric therefore remains asymptotically AdS up to order $1/N^2$.	
	
	\section{Transition amplitudes}
	
	We now discuss transition amplitudes and find somewhat different structure from the correlation function case. Nevertheless, we find that transition amplitudes admit an information metric in a certain sense. We study a quantum-mechanical setup that describes relevant features of transition amplitudes in quantum field theory. Consider a transition between states $\ket{\psi_f}, \ket{\psi_i}$ induced by unitary $U$. The transition amplitude is $\braket{\psi_f|U |\psi_i}$. In order to extract an information metric, we must expand about $D_B^2 = 0 $, but for $U \neq 1$, this does not necessarily occur when $\ket{\psi_i }= \ket{ \psi_f}$. 
	
	With density matrices $\rho_i = \ket{\psi_i}\bra{\psi_i}$ and $\rho_f = \ket{\psi_f} \bra{\psi_f}$, the following Bures distance contains the transition amplitude.
	\begin{equation}
	D_B(U^\dagger \rho_i U, \rho_f)^2 = 2 \left(1-\sqrt{U^{\dagger }\rho_i U  \rho_f} \right).
	\end{equation}
	Suppose $U=e^{-i\lambda H}$ for some dimensionless hermitian operator $H$. 	
	\begin{equation}
	D_B(U^\dagger \rho_i U, \rho_f)^2 = 2 \left(1-|\braket{\psi_f|U|\psi_i}|\right).
	\end{equation}
	Expanding in $\lambda$,
	\begin{align*}
	D_B(U^\dagger \rho_i U, \rho_f)^2 \approx 2 \bigg(1-&\bigg( |\braket{\psi_f|\psi_i}|^2
	-
	2\lambda \text{Im} \left( \braket{\psi_f|H|\psi_i} \braket{\psi_i|\psi_f} \right)
	\\
	&~~+
	\lambda^2 \left(|\braket{\psi_i|H|\psi_f}|^2 -\frac{1}{2}|\braket{\psi_i|H^2|\psi_f}|(\braket{\psi_i|\psi_f}+\braket{\psi_f|\psi_i})\right) \bigg)^{1/2}\bigg),
	\numberthis
	\end{align*}
	where Im$(a+ib) \equiv b$. If we choose $\ket{\psi_i} = \ket{\psi_f}$, the order $\lambda$ term above becomes zero. Expanding in $\lambda$ then gives
	\begin{equation}
	D_B(U^\dagger \rho_i U, \rho_f)^2 \approx \lambda^2
	\left( \braket{H^2} -\braket{H}^2 \right).
	\end{equation}	
	At $\ket{\psi_f} = \ket{\psi_i}$, the transition amplitude therefore admits the information metric at order $\mathcal{O}(\lambda^2)$
	\begin{equation}
	ds^2 = d\lambda^2 \left( \braket{H^2} -\braket{H}^2 
	\right).
	\label{TransitionAmpInfoMetric}
	\end{equation}
	For $\ket{\psi_i} \neq \ket{\psi_f}$, the states that have $D_B^2 = 0$ are $\ket{\psi_f} = U \ket{\psi_i}$. If we allow the states to vary independently, the full metric is the sum of \eqref{TransitionAmpInfoMetric} and the $\lambda = 0$ metric.
	
	To understand this discussion more explicitly, consider the following two qubit system.
	\begin{equation}
	\ket{\psi_i} = \begin{pmatrix} \cos\theta \\ \sin \theta \end{pmatrix} \otimes \begin{pmatrix} \cos\phi \\ \sin \phi  \end{pmatrix}, \ \ ~~~~~~~~~~~\ket{\psi_f} = \begin{pmatrix} \cos\theta^{\prime} \\ \sin \theta' \end{pmatrix} \otimes \begin{pmatrix} \cos\phi' \\ \sin \phi' \end{pmatrix}  . 
	\end{equation}
	For $\lambda =0 $,
	\begin{equation}\label{eqn:TwoQubitAmp}
	F(U^\dagger \rho_i U, \rho_f) =  \left|\cos{(\theta'-\theta)} \cos{(\phi' - \phi)}\right|^2.
	\end{equation}
	Expanding with $\theta'=\theta+ d\theta, ~ \phi'=\phi+ d\phi$, the information metric is
	\begin{equation}
	ds^2 = d\theta^2 + d\phi^2.
	\end{equation}
	Turning on the interaction $H = \sigma_z^1\otimes \sigma_z^2 $ gives
	\begin{equation}\label{eqn:TwoQubitAmpInt}
	D_B (U^\dagger \rho_i U, \rho_f)^2 \approx 2\left(1-\sqrt{\left(\cos{(\theta'-\theta)} \cos{(\phi' - \phi)}\right)^2  + \lambda^2 \left(\cos{(\theta' + \theta)} \cos{(\phi' + \phi)} \right)^2 } \right).
	\end{equation}
	Because the states and $H$ chosen were real, there is no $\mathcal{O}(\lambda)$ term above. Nevertheless, we still must check whether expanding about the point $\theta'=\theta+ d\theta, ~ \phi'=\phi+ d\phi$ gives a consistent information metric.
	\begin{align*}
		D_B(U^\dagger \rho_i U, \rho_f)^2 \approx  &\lambda^2 \left( 1- \cos^2{2\theta} \ \cos^2{2\phi} \right) \ 
		\\
		&+ d\phi \left( \lambda^2 \cos^2{2\theta} \ \sin{4\phi} \right) + d\theta \left(\lambda^2 \sin{4\theta} \cos^2{2 \phi}  \right) 
		- d\theta d\phi \left( \lambda^2 \sin{4\theta} \sin{4\phi} \right)
		\\	
		&+ d\theta^2 \left(1+ \frac{\lambda^2}{4} (-1+3\cos{4\theta}) \cos^2{2\phi} \right) 
		+ d\phi^2 \left( 1+\frac{\lambda^2}{4} (-1+3\cos{4\phi}) \cos^2{2\theta}   \right) 
		.
		\numberthis
	\end{align*}
	The $\mathcal{O}(d\theta,d\phi)$ terms are first order, proportional to $\lambda$, and non-zero. This was expected from the fact that $D_B^2 \neq 0$ at $\theta = \theta', \phi = \phi'$ once the interaction $H$ is included. We therefore have no meaningful information metric in $d\phi, d\theta$ at higher order in $\lambda$. The expansion of the information metric in $\lambda$ terminates,
		\begin{equation}
		ds^2 =  d\theta^2 + d\phi^2 + d\lambda^2 \left( 1- \cos^2{2\theta} ~ \cos^2{2\phi} \right),
		\end{equation}
		which agrees with \eqref{TransitionAmpInfoMetric} with $d\theta= d\phi = 0$.
		
		Applying this approach to the S-matrix may require some modification. Consider a unitary matrix $S$ written as $S = 1 + i T$. To obtain a form similar to \eqref{TransitionAmpInfoMetric}, we write
		\begin{equation}
		S = e^{-i H_S},
		\end{equation}
		where $H_S^\dagger = H_S$. Suppose $H_S$ can be expanded in a small parameter, $H_S =\sum_n \lambda^n H_S^{(n)}$. This leads to 
		\begin{equation}
		ds^2 = d\lambda^2 \left( \braket{(H_S^{(1)})^2}-\braket{H_S^{(1)}}^2  \right),
		\label{DiffSquares}
		\end{equation}
		of which \eqref{TransitionAmpInfoMetric} is a special case. \eqref{DiffSquares} corresponds to a transition amplitude with identical initial and final states.\footnote{Note that the information metric is nonzero only when these states are not eigenstates of $H_S^{(1)}$.} According to the quantum Cramer-Rao theorem, the error in estimating $\lambda$ from measuring the states is bounded from below by $\left( \braket{(H_S^{(1)})^2}-\braket{H_S^{(1)}}^2  \right)^{-1}$.
		
		\eqref{DiffSquares} is a completely general formula for transition amplitudes. It applies to transition amplitudes in position space as well as momentum space. Transition amplitudes have been studied in AdS/CFT \cite{BalasubramanianKLT98,BalasubramanianKL98,BalasubramanianGL99,Raju10,Raju12A,Raju12B,MeltzerS20,Sivaramakrishnan:2021srm}. The modular Hamiltonian $K_A=\log \rho_A$ generates unitary evolution within subsystem $A$, where states are defined on slices of constant modular time. \eqref{DiffSquares} therefore applies to transition amplitudes within the domain of dependence of $A$. It may be interesting to note that the quantity $\braket{K_A^2} - \braket{K_A}^2$ has been studied recently \cite{Verlinde:2019ade,Zurek:2020ukz,Banks:2021jwj}.
		
	\section{Future directions}
	We found that factorization, the OPE, and the $1/N$ expansion are encoded by information metric of correlators. It would be natural to flesh out this information geometry description: one can explore higher points, odd points, Lorentzian signature, operators with spin, twist operators, and so on. The interplay between $1/N$ corrections and quantum information ideas can be explored in this context. Special cases of our results yield the information metric of the two-point function in excited states, and of multitrace operators. Applications of more sophisticated ideas in quantum information geometry may produce new constraints on CFT data. We conclude by discussing a few directions in more detail.
	
	The information metric in principle encodes some or all of the same information as the original correlator. In this way, the information metric geometrizes the correlator in a seemingly novel fashion. It would be interesting if this description served as a useful organizing tool for CFT data. However, note that the information metric is derived from the normalized four-point function, which is a ratio of correlators and not a correlator itself. It would be interesting to understand this object better, though its appearance may suggest that quantities that are natural in information geometry are obscured in standard correlator language. In this spirit, it may be useful to understand what CFT features are encoded by the curvature scalar and tensors of the information metric. As multiple OPE channels are encoded by the information metric, can we impose crossing as a condition on the information geometry? If so, can this be used to derive new constraints on OPE data? It would be interesting to identify the information geometry of a single conformal block. On a more basic level, how do conformal transformations of the correlator act on the information metric?
	
	We have shown that cross terms in the metric signal a failure of factorization. They also represent non-trivial interplay between several different parts of the information geometry boundary, each of which is asymptotically AdS. It may be interesting to develop a better understanding of the full geometry. One could also ask whether higher-genus manifolds are allowed, and if so, what this would imply for the correlator. More modestly, can information manifolds of correlators have conjugate points, which appear in studies of complexity \cite{Balasubramanian:2019wgd,Balasubramanian:2021mxo}?
	
	The relationship between OPE data and complexity in holographic CFTs is not yet well-understood, though is natural to explore in light of recent work \cite{Chagnet:2021uvi,Flory:2020dja}. Ideas used in our work may be useful for studies of CFT complexity. Computing $1/N$ corrections to complexity may clarify its possible bulk dual \cite{Brown:2015bva}.
	
	In short, we have shown that information geometry provides a new representation of a large class of CFT correlators. While the usefulness of this representation remains to be seen, many new avenues are now open for exploration.
	
	\section{Acknowledgements}
	We thank Yuya Kusuki, David Meltzer, and Julio Parra-Martinez for discussions and comments on the draft. We thank Sinong Liu for initial collaboration on this work. AS thanks the Walter Burke Institute for Theoretical Physics for hospitality during the final stages of this work. The research of AS was supported by a National Science Foundation grant NSF-PHY/211673 and the College of Arts and Sciences of the University of Kentucky.
	
	\bibliographystyle{ssg}
	\bibliography{refs}
	
\end{document}